\documentclass[twocolumn,showpacs,preprintnumbers,amsmath,amssymb]{revtex4}

\usepackage{graphicx}
\usepackage{dcolumn}
\usepackage{bm}

\usepackage{exscale}
\usepackage{epsf,wrapfig,epsfig}

\newcommand{\vt}{Engineering Science and Mechanics and School of Biomedical Engineering and Sciences, Virginia Polytechnic Institute and State University, Mail Code 0219, Blacksburg, VA, 24061, USA}

\usepackage{color}

\makeatletter
\def\thefigure{\@arabic\c@figure}
\def\fps@figure{h, t}
\@addtoreset{equation}{section}

\makeatother

\begin{document}

\title{Separatrices and basins of stability from time series data}

\author{Martin Tanaka and Shane D. Ross}

\affiliation{\vt}

\date{\bf \normalsize This version: September 21, 2007}

\begin{abstract}
An approach is presented for identifying separatrices in phase space generated from noisy time series data sets representative of measured experimental data.  These separatrices are identified as ridges in the phase space distribution of finite-time Lyapunov exponents, i.e., Lagrangian coherent structures (LCS).  As opposed to previous approaches, the LCS is identified using only trajectories since no analytical or data-defined vector field is available.  The method is applied to a biological simulation in which the separatrix reveals a basin of stability.  These results suggest that the method will be a fruitful approach to time series analysis, particularly in cases where a limited number of trajectories are available as might be encountered in experiments.
\end{abstract}


\maketitle

Increasingly, dynamical systems of interest are defined not by analytical models, but by data from experiments or large-scale simulations, such as musculoskeletal biomechanics \cite{DiCu2000, Dingwell2006, EnGr2007} or geophysical fluid dynamics  \cite{Pierrehumbert1991, Pierrehumbert1991a, ShLeMa2005, Haynes2005}.

In many cases, researchers want to ascertain if deterministic chaos is present  \cite{Dingwell2006, Falconer2007}.  
This can be acheived by determining characteristic exponents that describe the sensitivity of the solution to initially close starting conditions.  One popular technique is to estimate the (maximum) Lyapunov exponent averaged over the sampled portion of phase space \citep{BeGaGiSt1980, BeGaGiSt1980a, Wolf1985, Eckmann1986, Rosenstein1993, KaSc2004}.  This method is well suited for analysis of time series data from experiments.

However, there is more information contained within the maximum Lyapunov exponents.  Instead of averaging the exponents over phase space to obtain a single scalar value, one can generate a maximum Lyapunov exponent field by considering how phase space expansion is distributed.
From this field, one can determine separatrices, co-dimension one boundaries in phase space separating qualitatively different kinds of motion.  

An example from biomechanics will demonstrate our point.
In biomechanics, a separatrix exists between standing and falling.  Standing with postural sway is a distinctly different type of motion than falling.  During standing, the body remains in the vicinity of an equilibrium position and may be characterized as dynamically stable.  Compare this motion to falling where the body rapidly diverges from the equilibrium position at an increasing velocity.  In falling, the body behaves unstably with respect to the upright vertical position.  

If we allow an experimental subject to take a step during fall recovery, another boundary will develop.  Now three states exist: standing, recovering from a fall with one step, and falling.  Each type of motion is divided from the other by a separatrix. Extending this theory, a phase space diagram with multiple fronts may be generated.

Similar to standing postural sway is the problem of maintaining torso stability.  Torso stability is necessary to avoid large deformations in the lumbar spine that may result in low back injury and pain.  In this Rapid Communication, a method will be developed to locate the basin of stability that separates stable, injury free torso sway from unstable, potentially injurious motion.

The computation of finite-time Lyapunov exponents from experimental data has been used before in musculoskeletal biomechanics, particularly to quantify local dynamic stability during locomotion \citep{DiCu2000}. We demonstrate that the phase space distribution of finite-time Lyapunov exponents can also provide information.  This information includes the boundaries, or separatrices, between qualitatively different states, as described above for the biomechanics example.

The separatrices are found as Lagrangian coherent structures (LCS), borrowing a term from the fluid mechanics \citep{HaYu2000}, and are defined as the ridges of the finite-time Lyapunov exponent field \citep{ShLeMa2005}.  These structures indicate the location of the separatrix demarking the boundary between qualitatively different kinds of motion.   LCS have previously been used to analyze dynamical systems defined by fluid flow fields from data \citep{Haller2002, WaHaBaTa2003} and analytical biochemical models \citep{AlHaSoLa2006} but have not been applied to biomechanical data or {any time series data obtained from experiments}. 


The goal of this Rapid Communication is to demonstrate that LCS can be used to analyze experimental time series data where only individual trajectories are available.  LCS are robust with respect to noise \citep{Haller2001, Haller2002} making them even more attractive for use in experimental data analysis where noise sensitivity is an important issue \citep{Casdagli1991, ElTu1995, FrSa2001}.  


This Rapid Communication is organized as follows: first, we discuss the method for estimating the maximum finite-time Lyapunov exponent (FTLE) associated with a point in phase space.
Second, we demonstrate how the phase space distribution of maximum FTLE can be obtained using only trajectories.  Third, using a model of human postural control, we demonstrate how ridges can be found in the phase space distribution of the maximum FTLE.  These ridges are separatrices partitioning the phase space into regions of distinct behavior.

In the following discussion, we borrow ideas developed in fluid mechanics \citep{HaYu2000, Haller2000, Haller2001, Haller2001a, Haller2002, ShLeMa2005} and with some modifications apply them to a new context, a situation where only trajectories are available and not the vector field itself.

{\it Sensitivity analysis and finite-time Lyapunov exponents.}---Suppose we are given a reference trajectory $x(t)$ going from $x_0$ at time $t_0$ to $x_1$ at time $t_1$.  We assume the trajectory evolves under the dynamical equations of a time-independent (autonomous) system
\begin{equation}\label{system_eqn1}
\dot x= f(x),~~~x \in \mathbb{R}^n.
\end{equation} 
This equation describes a flow field or vector field.
The sensitivity of the reference trajectory is discussed below.

Let trajectories of the system (\ref{system_eqn1}) with $x(t_0)=x_0$ be denoted by $\phi (t,t_0)$.  
In other words, $\phi (t,t_0) :x(t_0) \mapsto x(t)$ denotes the {flow map} of the dynamical system (\ref{system_eqn1}), mapping particles from their initial location at time $t_0$ to their location at time $t$.  
For our purposes, we will denote the flow map as $\phi (t,t_0;x_0)$ or simply $\phi (t;x_0)$ so the dependence on the initial condition $x(t_0)=x_0$ is made clear.

Consider a second trajectory that starts slightly away from the reference trajectory $x(t)$, i.e., starts from the perturbed initial vector $x_0+\delta x_0$ at time $t_0$.  As the trajectories evolve, the vector displacement (or perturbation vector)
\begin{equation}\label{delta_x_t}
\delta x(t)=\phi (t; x_0+\delta x_0)-\phi (t; x_0)
\end{equation}
will also evolve.  
For our purposes, the ``second trajectory'' might be the result of another experimental trial or another portion of the same trajectory separated in time by a minimum threshold value.  We discuss this further below.

The linear relationship between small initial perturbations and perturbations at some time $t$ is   
\begin{equation}\label{STM1}
\delta  x(t)=\Phi (t,t_0)
\delta  x_0.
\end{equation}
where 
$\Phi (t,t_0)={\frac{\partial \phi (t; x_0)}{\partial x_0}}$ is the  {state transition matrix} (also known as the {fundamental matrix}).
The state transition matrix 
can be viewed as a deformation gradient.  If an (infinitesimal) $n$-dimensional spherical blob of particles is placed about the reference trajectory, then after a duration $T=t-t_0$, the blob will have expanded in some directions and compressed in others to form an $n$-dimensional ellipsoid.  
The matrix $\Phi (t,t_0)$ contains information about this expansion and contraction {as well as the rotation} of the initial blob of particles, due to the locally deforming nature of the flow.  

Suppose there exists a state transition matrix over some interval, $\Phi (t,t_0)$.  
The size of the final perturbation at time $t$ is given by
\begin{equation}\label{STM2}
\| \delta  x(t) \|^2 
= \delta  x_0^{\ast} [\Phi (t,t_0)^{\ast} \Phi (t,t_0)] \delta  x_0,
\end{equation}
where $\| \cdot \|$ is the vector norm on $ \mathbb{R}^n$, $A^{\ast}$ denotes the transpose of the matrix $A$, and the perturbations are considered as column vectors.
The symmetric matrix
\begin{equation}\label{Cauchy-Green}
C = \Phi (t,t_0)^{\ast} \Phi (t,t_0),
\end{equation}
is the finite-time right Cauchy-Green deformation tensor \citep{ShLeMa2005}.  
The matrix $C$ is a rotation-independent measure of deformation; it gives the square of the local change in distances due to deformation \citep{TrNo2004, Fung1993}.
Since $C$ is a symmetric, positive definite matrix, it has $n$ real, positive eigenvalues \citep{Strang1988, LeShMa2007}.

One can associate with point $x_0$ a maximum {finite-time Lyapunov exponent}, given by
\begin{equation}\label{maxFTLE}
\sigma_{1}(x_0) = \frac{1}{T} \ln \sqrt{\lambda_\text{max}(C)},
\end{equation}
where $T = t-t_0$ is the finite duration over which expansion is measured and ${\lambda_\text{max}(C)}$ is the 
maximum eigenvalue of $C$ with the corresponding (normalized) eigenvector $\hat{e}_1(t_0)$.  In other words, if $\delta x_0$ is along $\hat{e}_1(t_0)$ at time $t_0$, then maximum stretching occurs over the time $T$ and the length of the perturbation vector becomes 
\begin{equation}\label{maxFTLE_growth}
\| \delta x (t) \| = e^{\sigma_{1}(x_0) T} \| \delta x_0 \|,
\end{equation}
where $t=t_0 +T$ \citep{ShLeMa2005}.

{\it Computing the maximum FTLE field when only individual trajectories are available.}---If we assume that the direction of maximum expansion dominates the dynamics of perturbations in arbitrary directions \citep{Rosenstein1993}, we can approximate the maximum FTLE field.
Under this assumption, we assume that (\ref{maxFTLE_growth}) holds for {all } perturbation vectors, regardless of initial phase space direction.  This will result in a conservative approximation to the actual value when the arbitrary vector direction is not aligned with the direction of maximum expansion.  
As our concern is now focused on the maximum FTLE, we will refer to the maximum FTLE as simply the FTLE hereafter.



\begin{figure}[h,t]
\begin{center}
\includegraphics[width=0.5\textwidth]{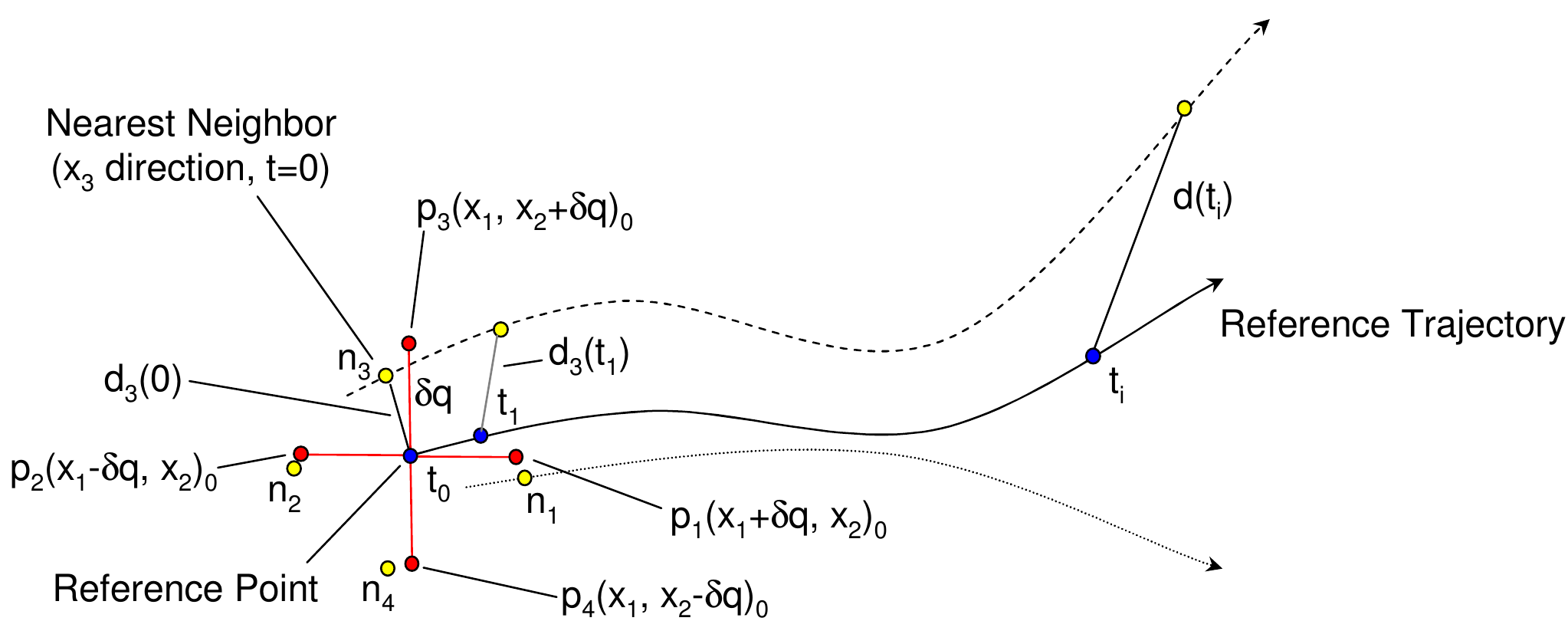}
\end{center}
\caption{\label{FTLE-nearest-neighbor}{\footnotesize
Estimating the maximum FTLE by averaging over the growth of perturbation vectors in multiple phase space directions.  We make the assumption that the maximum FTLE dominates the evolution of the perturbation vectors.}}
\end{figure}

The FTLE is estimated as the rate of separation of neighboring trajectories.  In order to understand how this is determined from experimental data, consider the reference trajectory shown in Figure \ref{FTLE-nearest-neighbor}.
With a reference point established, a target location $(p_1)$ is identified that is a perturbation distance $\delta q$ from the reference point in phase space.  The data point closest to $p_1$ on another trajectory is then found, $n_1$.  The other trajectory can be from either a different experimental trial or another portion of the same trial separated by a sufficient amount of time to avoid a correlation with the reference point.  This process is repeated for other directions.  Using this method, 2$n$ neighbors are found for each reference point corresponding to positive and negative directions of each dimension of state space.  If multiple nearest neighbors are considered, perturbations are sampled in multiple phase space directions which increases the likelihood of a separation lying in the direction of maximum expansion. 

Each point in the data set is sequentially evaluated by considering it to be a reference point.  
The FTLE is calculated for each pair, but unlike previous analyses that averaged the FTLE over time and phase space \citep{EnGr2007,  Rosenstein1993, Dingwell2006, Wolf1985}, the exponent value over a finite time $T$ is associated with a {phase space location} midway between the pair of points.  In this way, the phase space distribution of the FTLE is developed.  The FTLE field is generated by placing an $n$-dimensional grid over the phase space and using the distribution of FTLE to generate an $n$-dimensional surface.  At each phase space location, $x$, the height of the surface is the scalar value $\sigma_1(x)$ from equation(\ref{maxFTLE}).  

The choice of $\delta q$ influences the FTLE field.  The perturbation distance $\delta q$ is a coarse-graining parameter selected to be large enough to overcome system noise and small enough to reveal local features of the FTLE field.  

\begin{figure}[h,t]
\begin{center}
\includegraphics[width=0.40\textwidth]{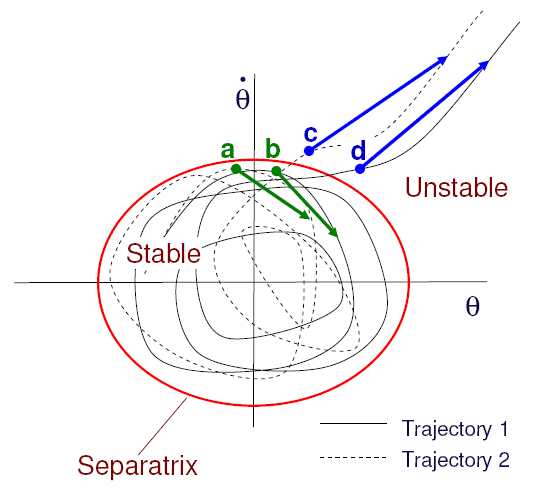}\\
\end{center}
\caption{\label{LCS-schematic}{\footnotesize
The divergence of two points of opposite sides of a separatrix is larger than the divergence of points on the same side. This generates a ridge in the FTLE field at the separatrix.}}
\end{figure}

Figure \ref{LCS-schematic} is used to explain why ridges develop in the FTLE field at the separatrices.  In this schematic representation, divergence of two points (a and b) within the stable region and two points (c and d) in the unstable region are compared.  As the points a and b evolve in time, their trajectories only slightly diverge resulting in a small FTLE.  This value forms a data point on the FTLE field midway between a and b.  Similarly, points c and d also diverge slightly resulting in a small FTLE and another data point for generating the FTLE field.  However, b and c on opposites sides of the separatrix diverge greatly even over short times.  This results in a large FTLE that forms a data point on the FTLE field at the approximate location of the separatrix. As more data points are entered a ''volcano shaped'' ridge forms revealing the separatrix.  

\begin{figure}[!b]
\begin{center}
\includegraphics[width=0.40\textwidth]{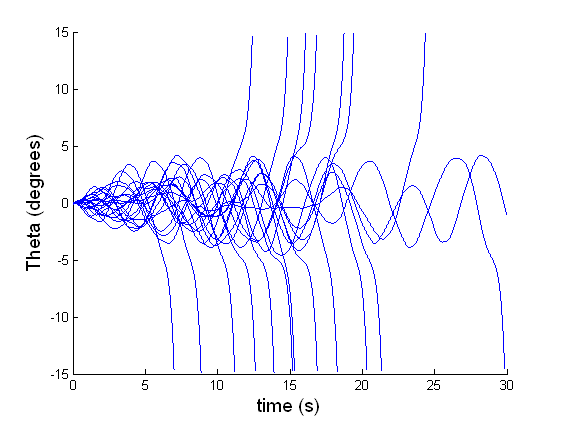}
\end{center}
\caption{\label{time_series}{\footnotesize
The time-series data analyzed came from several simulated experimental trials.}}
\end{figure}

{\it Numerically simulated experiments with noise.}---We have implemented the above approach for a model of human postural control: the inverted pendulum with limited gain proportional control. A simulation with 20 independent trials was generated from the model (Figure \ref{time_series}).  In the simulation, the system begins at equilibrium and is perturbed with Gaussian random force perturbations.  These forces generate movement that is attenuated by a controller.  As a result, the system is able to maintain stability for a period of time before an unrecoverable sequence of perturbations causes the system to become unstable.  Therefore, an important aspect of this simulation is that it spans the space of possible states.  This is important because it is not possible to experimentally determine the edge of the basin of stability unless data is available on both sides of the separatrix.



 
Phase space reconstruction is considered a preprocessing step in our method.  For the biomechanical system used to demonstrate the method, we do not use time-delay reconstruction.  Since the system under study is a mechanical system, we take the measured coordinate ($q$) and numerically construct the time derivative ($\dot q$), taking the 2D space of $x=(q,\dot q)$ as the reconstructed phase space.  However, we note that the method of using an FTLE field to find separatrices is not tied to any particular means of phase space reconstruction.
 
\begin{figure}[!t]
\begin{center}
\includegraphics[width=0.45\textwidth]{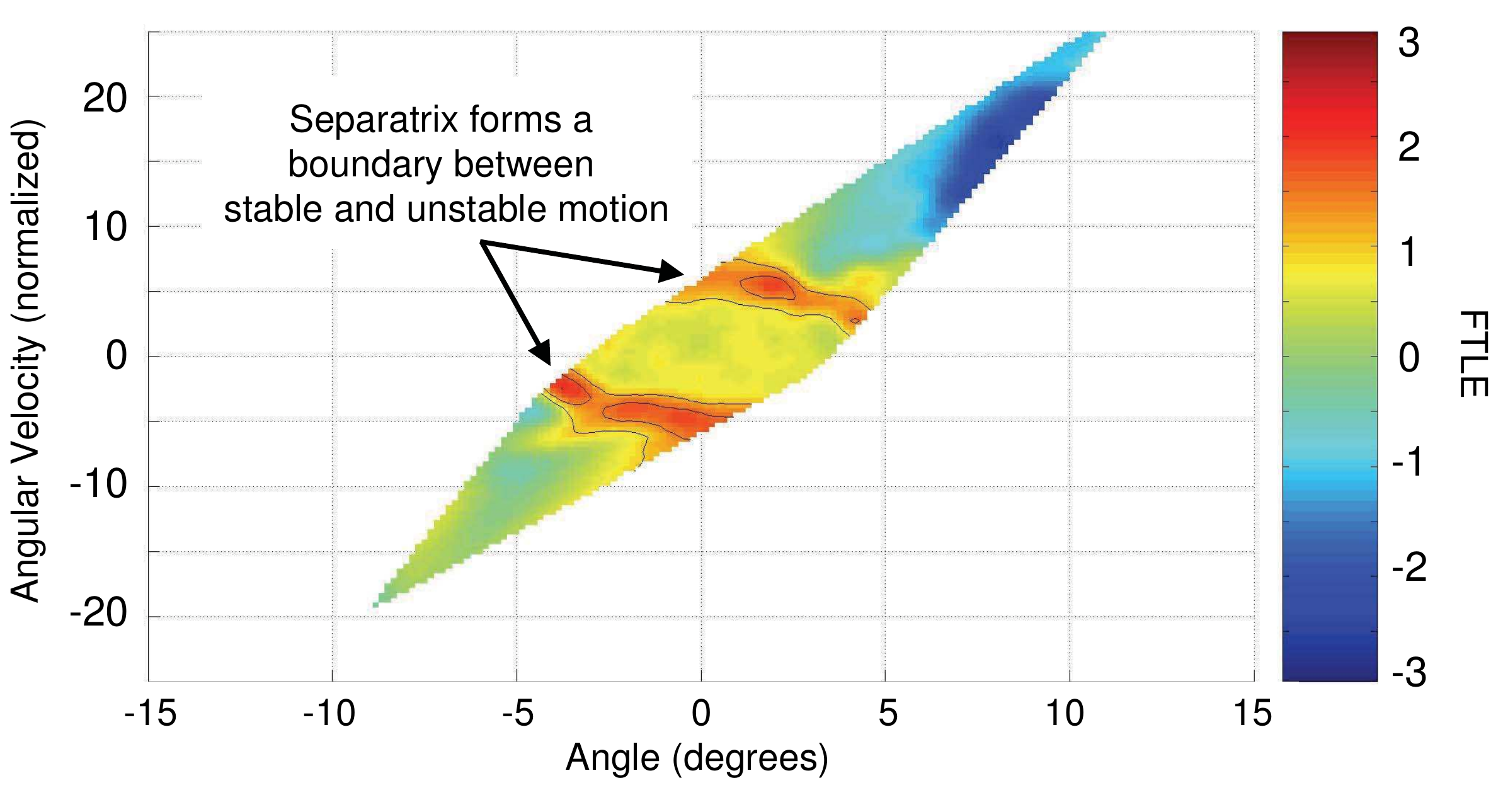}
\end{center}
\caption{\label{LCS-pendulum-standing}{\footnotesize
A clearly observable separatrix forms the boundary of the basin of stability around an equilibrium region.}}
\end{figure}

The resulting FTLE field and separatrix for an evolution time of  $T=1.2$ seconds are shown (Fig.~\ref{LCS-pendulum-standing}).  This value of $T$ approximately corresponds to the characteristic timescale of evolve in the unstable region. 


Two ridges in the distribution of maximum FTLE are discernible. 
As shown elsewhere \citep{ShLeMa2005}, ridges in the FTLE field reveal the location of partial transport barriers or separatrices in the phase space, providing boundaries which partition the space into regions of different behaviors.
In this case, the separatrix forms a boundary between the region of stable postural sway (around the origin) and falling motion (beyond the boundary).  The characteristics of the stable region and its boundary location in this biological example depend on the accuracy of a number of neurological sensory systems, the feedback gain associated with core muscle strength, and the time delay of the postural control system.

Although 20 trials were evaluated for this simulation, a separate simulation indicates that LCS structures may be identified with as little as two time series using this method.  However, smaller trial numbers will tend to be more affected by random noise than larger sets because the mean value of random noise approaches zero as the number of trials increases.

{\it Conclusions and future directions.}---We have shown in this Rapid Communication that even without an analytically defined or data defined vector field, we are able to identify separatrices using the LCS method.  A dynamic model of human postural control driven by noise was used to generate a time series similar to experimental data.  When we applied the LCS method to the simulated data, a separatrix was revealed which formed the boundary of a basin of stability around an equilibrium location.  As a result, we believe this method provides a fruitful approach for extracting information from noisy experimental data on boundaries between qualitatively different kinds of motion.

We note that the phase space averaged FTLE for any time $T$ can be obtained by computing the average of the FTLE field over the sampled region of phase space.  This provides the link between the current method and previous methods for finding an averaged Lyapunov exponent. 

In forthcoming work, we will demonstrate the method on higher dimensional data and actual experimental data.  We will also consider another method for constructing an FTLE field, one which estimates the state transition matrix $\Phi(t,t_0)$.  We believe this may be a more accurate method for obtaining the FTLE field over short evolution times, $T$.

{\small
\renewcommand\refname{References}

}
\end{document}